\documentclass{aa}  

\usepackage{graphicx}
\usepackage{txfonts}
\usepackage{hyperref}
\usepackage{todonotes}

\newcommand{\parsec}{{\sc parsec}}
\newcommand{\sevn}{{\sc sevn}}

\definecolor{seagreen}{rgb}{0.190, 0.525, 0.361}

\newcommand{\orcit}[1]{\protect\href{https://orcid.org/#1}{\protect\includegraphics[width=8pt]{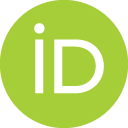}}}

\begin{document}

   \title{The impact of envelope binding energies on the merger rate density of binary compact objects}

   \titlerunning{Envelope binding energies and merger rate densities}
   
   \subtitle{}

   \author{Cecilia Sgalletta\orcit{0009-0003-7951-4820} \inst{1,2,3,4}\thanks{\href{mailto:cecilia.sgalletta@uni-heidelberg.de}{cecilia.sgalletta@uni-heidelberg.de}},
   Guglielmo Costa\orcit{0000-0002-6213-6988} \inst{5,6},
   Giuliano Iorio\orcit{0000-0003-0293-503X} \inst{7},
   Kendall Shepherd \orcit{0000-0001-5231-0631} \inst{2,6},
   \\Francesco Addari \orcit{0000-0002-3867-9966} \inst{2},
   Alessandro A. Trani \orcit{0000-0001-5371-3432} \inst{3,8,9},
   Michela Mapelli \orcit{0000-0001-8799-2548} \inst{1,5,10,11},
   Ugo N. Di Carlo\orcit{0000-0003-2654-5239} \inst{2},
   Andrea~Lapi\orcit{0000-0002-4882-1735} \inst{2,12,13},
   Alessandro~Bressan\orcit{0000-0002-7922-8440} \inst{2},
   Mario Spera\orcit{0000-0003-0930-6930}\inst{2,3,4}\thanks{\href{mailto:mspera@sissa.it}{mspera@sissa.it}},
          }

    \institute{
    Universit\"at Heidelberg, Zentrum f\"ur Astronomie (ZAH), Institut f\"ur Theoretische Astrophysik, Albert-Ueberle-Str. 2, 69120, Heidelberg, Germany
    \and
    SISSA, via Bonomea 365, I--34136 Trieste, Italy
    \and
    National Institute for Nuclear Physics – INFN, Sezione di Trieste, I--34127 Trieste, Italy
    \and
    Istituto Nazionale di Astrofisica – Osservatorio Astronomico di Roma, Via Frascati 33, I--00040, Monteporzio Catone, Italy
    \and
   Physics and Astronomy Department Galileo Galilei, University of Padova, Vicolo dell'Osservatorio 3, I--35122, Padova, Italy 
   \and
    INAF-Osservatorio Astronomico di Padova, Vicolo dell’Osservatorio 5, Padova, Italy
    \and
    Departament de F\'isica Qu\`antica i Astrof\'isica, Institut de Ci\`encies del Cosmos, Universitat de Barcelona, Mart\'i i Franqu\`es 1, E-08028 Barcelona, Spain 
    \and
    Niels Bohr International Academy, Niels Bohr Institute, Blegdamsvej 17, 2100 Copenhagen, Denmark
    \and
    Departamento de Astronomía, Facultad Ciencias Físicas y Matemáticas, Universidad de Concepción, Avenida Esteban Iturra, Casilla 160-C, Concepción, 4030000, Chile
    \and
   INFN - Padova, Via Marzolo 8, I--35131 Padova, Italy
    \and
    Universit\"at Heidelberg, Interdisziplin\"ares Zentrum f\"ur Wissenschaftliches Rechnen, Heidelberg, Germany
    \and
    Institute for Fundamental Physics of the Universe – IFPU, Via Beirut 2, I--34014 Trieste, Italy
    \and
    Istituto di Radioastronomia – INAF/IRA, Via Piero Gobetti 101, I--40129 Bologna, Italy\\}

   % \date{Received September 15, 1996; accepted March 16, 1997}

   \abstract{The common envelope (CE) phase plays a key role in the formation of binary compact object systems. Its final outcome strongly depends on the envelope binding energy, but this quantity is often estimated using fitting formulas that are not fully consistent with the underlying stellar evolution models adopted in population-synthesis codes. Here, we investigate envelope binding energies across the most extensive stellar grid considered to date. Our stellar tracks, evolved with \parsec{} v2.0, include hydrogen (H) -rich stars with metallicities ranging from $Z = 10^{-11}$ (Population III stars) to $Z = 0.03$, and initial masses between 2 and 2000 M$_\odot$, as well as pure-helium stars with masses from 0.36 to 350 M$_\odot$. We examine the sensitivity of the envelope binding energies to the selected core-envelope boundary definition and to different internal energy source contributions. 
   For H-rich stars, we find that internal energy sources can alter the envelope binding energy by more than an order of magnitude, whereas the core boundary criteria play a secondary role. In contrast, for pure helium stars, the core-boundary criterion becomes the dominant factor.
   The envelope binding energies derived from different stellar tracks can show deviations of several orders of magnitude, with larger differences for more massive stars and higher metallicities. 
   Finally, by implementing our new envelope binding energy prescriptions into the binary population synthesis code \sevn{}, we show that the predicted merger rate densities of compact binaries can differ by more than an order of magnitude compared to previous models. Our results highlight the importance of using envelope binding energies that are consistent with the underlying stellar evolution models and caution against extrapolating empirical fits beyond the considered parameter space.
   }

   \authorrunning{C. Sgalletta et al.}

   \maketitle
%
%-------------------------------------------------------------------

\section{Introduction}

The common envelope (CE) phase is one of the most prominent, yet poorly understood processes in binary stellar evolution. It denotes a rapid phase of evolution during which the two cores of the stars orbit within a shared envelope \citep{Paczynski1976, webbink1984, Iben1993, Podsiadlowski2001, Ivanova2013a}. The drag forces cause the orbit to shrink with exceptional efficiency, resulting in a merger of the two objects, or in a post-CE system significantly more compact than its original state.

The CE process is commonly invoked to explain several observed types of binary objects. Remarkable examples include cataclysmic variables \citep{Meyer1979}, type Ia supernovae \citep{Iben1984, webbink1984, Han2004}, binary pulsars \citep{vandenHeuvel1994, Vignagomez2020} and close white dwarf binaries \citep{Paczynski1976, Han2002, Han2003, toonen2013}. Such systems typically require a phase of efficient spiral-in, reducing the orbital separation by several orders of magnitude \citep{roepke2023}. 
Moreover, the CE phase holds a key role in the formation and merger of binary compact objects
\citep{Giacobbo2018, Kruckow2018, Breivik2020, Broekgaarden2021, Bavera2021, Romangarza2021, Mapelli2021, Belczynski2020, iorio2023,  Sgalletta2024}.

The CE process poses a tough computational challenge due to the large ranges involved both in time and space \citep{Ivanova2013b, MacLeod2017, Fragos2019, ivanova2020}. For this reason, hydrodynamical simulations still struggle to model the full CE evolution or fail to incorporate some physical ingredients, such as recombination energy, nuclear reactions that may take place within the dense accreted material, neutrino cooling, radiative and convective transport, jet formation \citep[see, e.g.,][ for a review]{roepke2023}.
Moreover, there is still no consensus in the literature regarding the final outcomes of a CE event, with many hydrodynamical simulations failing to achieve complete ejection of the envelope \citep{Taam2000, Passy2012, Ricker2012, Ohlmann2016, Iaconi2017, Moreno2022}. Notably, \citet{Nandez2015} find that including the contribution from the recombination energy allows to successfully unbind of the envelope.

Binary population synthesis codes usually adopt simplified analytic approaches because of their easier implementation and computational efficiency. Usually these are parametric models based on general considerations of energy \citep{webbink1984} and angular momentum \citep{Nelemans2000, nelemans2005} conservation, although also new formalisms have recently been proposed \citep{Trani2022, Hirai2022, diStefano2023}.
The $(\alpha\lambda)-$formalism \citep{vandenheuvel1976, webbink1984} is one of the most commonly adopted approaches in population synthesis codes. This framework is based on an energy balance equation in which the two components at play are the binding energy of the envelope (parametrized by $\lambda$) and the orbital energy  of the system. A second parameter, $\alpha$, governs the efficiency of the energy transfer mechanism.

It has been clear for a long time that the $(\alpha{}\lambda)-$formalism fails to capture the physics of CE evolution, namely all the sources and sinks of energy that are at play \citep{Ivanova2013b,Fragos2019}. However, population-synthesis codes still use this formalism because of its intrinsic simplicity,  and "hide" into $\alpha{}$ and/or $\lambda{}$  all the uncertainties about missing physics. For instance, this is the reason why several authors adopt  values of $\alpha>1$ \citep{Mapelli2018,Fragos2019,Hirai2022}. 
Furthermore, $\alpha$ may not be constant across different systems, but there may be a discrepancy between low-mass and high-mass donors \citep{Politano2007, deMarco2011, wilson2022}. 

The parameter $\lambda$ encapsulates the structural properties of a star's envelope, and is intrinsically linked to its stellar mass, metallicity, and evolutionary phase. $\lambda$ is typically derived from stellar evolution models. However, its determination can vary considerably due to differences among these models. Most population-synthesis codes adopt values of $\lambda$ that are not calculated directly from the adopted evolutionary tracks. This introduces a methodological inconsistency between the adopted stellar evolution models and the envelope binding energy prescriptions.
Additionally, there are still open questions regarding the correct evaluation of $\lambda$. The primary uncertainties arise from the criterion used to define the core-envelope boundary and from the contributing energy sources that should be considered in the evaluation of the binding energy \citep{Ivanova2013b, Wang2016, Kruckow2016}. 
The uncertainties on $\alpha$ and $\lambda$, in turn, reflect on the demographics of binary compact objects, producing variations of orders of magnitude in the resulting merger rates
\citep{Belczynski2002, Neijssel2019, Broekgaarden2021, Broekgaarden2022, iorio2023, Sgalletta2023, Boesky2024, Sgalletta2024}.

\cite{Kruckow2016} and \cite{klencki2021} conducted detailed studies of the envelope ejection conditions by massive stars. They showed that envelope binding energies are highly sensitive to the adopted $\lambda$ prescriptions, with different choices leading to large variations in the inferred values. Furthermore, while $\lambda$ parametrizes the binding energy of the whole stellar envelope, the outcome of the CE phase may be more closely linked to the properties of the outer convective envelope, which is expected to dominate the initial dynamical inspiral and therefore regulate the efficiency of envelope ejection \citep{klencki2021, Hirai2022, Picker2024}.

Here, we present envelope binding energies for a wide set of stars, significantly extending the range of stellar masses and metallicities explored in previous studies.
The set has been simulated with the PAdova and TRieste Stellar Evolution Code, \textsc{parsec} v2.0 \citep[][]{bressan2012, costa2019a, Costa2021, Nguyen2022, Costa2025, Shepherd2025}. Notably, we expand this analysis to include pure-He stars, addressing previously unconstrained $\lambda$ parameters. Moreover, we include the new sets of binding energies within the population synthesis code \textsc{sevn} \citep{Spera2017, Spera2019, Mapelli2020b, iorio2023} and we employ the new formalism to test the cosmic merger rate densities for binary compact objects. In agreement with recent works, our results emphasize the importance of consistent envelope binding energies. Moreover, our results hint towards a functional dependence of the $\alpha$ parameter. 

In Section \ref{sec:method} we describe the methodology adopted in this work. In Section \ref{sec:results} we present the envelope binding energies, emphasizing the importance of convection. Furthermore, we describe their implementation in \textsc{sevn} and show how they compare with existing models. We discuss our findings in Section \ref{sec:discussion} and draw our conclusions in Section \ref{sec:summary}.

\section{Method} \label{sec:method}

\subsection{\textsc{parsec}}

We use stellar tracks computed with the stellar evolution code \parsec{} v2.0 \footnote{\href{https://stev.oapd.inaf.it/PARSEC/index.html}{https://stev.oapd.inaf.it/PARSEC/index.html}} \citep[][]{bressan2012, costa2019a, Costa2021, Nguyen2022, Costa2025, Shepherd2025}. 
We use two sets of tracks from \citet{Costa2025}. In the first set, we start from the hydrogen main sequence (hereafter H tracks). In the second, we start from the helium main sequence, i.e. our tracks have hydrogen fraction $X_{\rm H}=0$ since the beginning (hereafter, pure-He tracks). 
The H tracks have metallicity values $Z=10^{-11}$, $10^{-6}$, $0.0001$, $0.001$, $0.002$, $0.004$, $0.006$, $0.008$, $0.01$, $0.014$, $0.017$, $0.02$ and $0.03$. The zero-age main sequence (ZAMS) masses $M_{\rm ZAMS}$ range from $2.2$ to $600$~M$_{\odot}$ for $Z\geq 0.001$, $M_{\rm ZAMS}$ reaches up to $2000$~M$_{\odot}$ for lower metallicities.
The models assume overshooting on top of convective cores,
allowing for the penetration of convective elements into stable regions under the "ballistic" scheme described by \citet{Bressan1981}. Within this scheme the overshooting length - corresponding to the maximum distance traveled by the overshooting bubble across the unstable border - is roughly 0.25 pressure
scale heights. Additionally, all the tracks have been evolved assuming no rotation. 
The tracks include, for each evolutionary timestep, detailed spatial grids of the main stellar properties. To the purposes of our work, we are interested on the radial distance ($r$), gravitational energy ($E_{\rm g}$), internal energy ($U$), enthalpy ($E_{\rm H}$), luminosity ($L$), hydrogen and helium concentrations ($X_{\rm H}$ and $X_{\rm He}$, respectively). 
For stars with $2 \leq M_{\rm ZAMS} / $M$_\odot \leq 9$, we consider their evolution until the early asymptotic giant branch phase. Massive stars' evolution ($M_{\rm ZAMS} > 9$~M$_\odot$) is followed until either the start of core oxygen burning or the pair instability supernova regime. We refer to \citet{Costa2025} for further details on the stellar tracks.

For the pure-He tracks, the zero-age He main sequence mass ($M_{\rm He}$) ranges from $0.36$ to $350$M~$_{\odot}$ and metallicities: $Z=10^{-6}$, $0.0001$, $0.0002$, $0.0005$, $0.001$, $0.002$, $0.004$, $0.006$, $0.008$, $0.01$, $0.02$, $0.03$ and $0.05$.
We refer to \cite{Costa2025} for additional details on the tracks.

\subsection{Envelope binding energy}\label{sec:ebind}

The most general expression of the gravitational binding energy of the envelope is
\begin{equation} \label{eq:eg}
    E_{\rm G} = - \int^{M}_{M_{\rm core}} \frac{G m}{r(m)} dm
\end{equation}
where $M$ is the total stellar mass, $M_{\rm core}$ is the core mass, $m$ is the local mass coordinate.
The quantity $r(m)$ is the radius from the center at mass coordinate $m$. 
The internal energy $U$ (including both thermal and recombination energies) may also contribute lowering the binding energy of the envelope \citep{dewi2000}, therefore we can define $E_{\rm B}$ as
\begin{equation} \label{eq:eb}
    E_{\rm B} = \int^{M}_{M_{\rm core}} \left[ - \frac{G m}{r(m)} + U(m) \right]  dm.
\end{equation}
Finally, \cite{ivanova2011a} first suggested the pressure work term as an additional energy source, contributing to the overall binding energy of the envelope ($E_{\rm H}$):
\begin{equation}\label{eq:eh}
    E_{\rm H} = \int^{M}_{M_{\rm core}} \left[ - \frac{G m}{r(m)} + U(m) + \frac{P(m)}{\rho(m)}\right]  dm,
\end{equation}
where the term $U+ P/\rho$ is known as the specific enthalpy. 

The boundary between the core and the envelope is a key quantity for the computation of these integrals. A number of criteria have been proposed \citep[see e.g.][]{Ivanova2013b}. We define the helium core of a star as the hydrogen-depleted region; where the fraction of hydrogen ($X_{\rm H}$) is lower than a fixed threshold $X_{\rm H, 0}$. The commonly adopted value in the literature is $X_{\rm H, 0} \sim 0.1$ \citep{Xu2010, ivanova2011b, klencki2021, Marchant2021}. On the other hand, the \parsec{} code defines the core as the region  $X_{\rm H, 0}$ reaches $10^{-3}$ by default.
We keep $X_{\rm H, 0}$ as a free parameter and we vary its value in our models in order to explore its effect. 
We evaluate $E_{\rm G}$, $E_{\rm B}$ and $E_{\rm H}$ through eq. \ref{eq:eg}, \ref{eq:eb} and \ref{eq:eh} at each timestep of the stellar evolution where $M_{\rm core} \neq 0$. We repeat this procedure for every stellar track.

Whether or not a pure-He star can initiate a second unstable mass transfer episode is still unclear. According to the prescription by \citet{hurley2002}, mass transfer from pure-He stars becomes dynamically unstable if the donor-to-accretor mass ratio is larger than a critical mass ratio $q_{\rm crit}=3$ for He main sequence stars, and $q_{\rm crit}\sim{1}$ (or even lower) in more advanced evolutionary stages. Other works assume mass transfer from pure-He stars to always be dynamically stable \citep{Tauris2015, Tauris2017, Neijssel2019}. In particular, \citet{Vignagomez2018} find that this condition is required in their models in order to explain the population of Galactic binary neutron stars. Nevertheless, \citet{Sgalletta2023} show that this assumption is not necessary adopting \textsc{sevn} and the \textsc{parsec} stellar tracks. 
We also study the envelope binding energies for pure-He stellar tracks. In this case, $E_{\rm G}$, $E_{\rm B}$ and $E_{\rm H}$ are computed with equations \ref{eq:eg}, \ref{eq:eb} and \ref{eq:eh}, respectively. However, in this case, the lower limit of integration is the carbon-oxygen (CO) core mass. The latter is defined in turn, as the He-depleted region: $X_{\rm He} < X_{\rm He, 0}$, with $X_{\rm He, 0}$ a concentration threshold similar to the one previously defined for hydrogen.

Here, we study how the binding energies vary for different parameters in our models. We test multiple core separation prescriptions, varying $X_{\rm H, 0}=10^{-1}, 10^{-3}, 10^{-7}$ and $X_{\rm He, 0}=10^{-1}, 10^{-3}, 10^{-7}$. Moreover, we test how the different definitions $E_{\rm G}$ (eq. \ref{eq:eg}), $E_{\rm B}$ (eq. \ref{eq:eb}), and $E_{\rm H}$ (eq. \ref{eq:eh}), impact the results. 

\subsection{Stellar EVolution for N-body code (\textsc{sevn})} \label{sec:sevn}
\textsc{sevn}\footnote{  In this work, we use the \textsc{sevn} version V 2.16 (commit \href{https://gitlab.com/sevncodes/sevn/-/tree/8af02cc3706b132609cb3b7fb3ca029c8629bfb7}{8af02cc3}). \textsc{sevn} is publicly available at the gitlab repository \url{https://gitlab.com/sevncodes/sevn}}  is a fast and versatile binary population synthesis code \citep{Spera2017, Spera2019, Mapelli2020, iorio2023}. 
It interpolates the main stellar properties (masses, radii, luminosities, etc) on the fly from a precomputed set of stellar tracks, and accounts for  binary-evolution processes through analytic and semi-analytic formulas.

For the interpolation of stellar properties, \sevn{} includes seven tables: evolutionary time, total stellar mass, He-core mass, CO-core mass, stellar radius, bolometric luminosity, and the stellar phase. Other  properties can be included by adding further optional tables 
\citep[see ][for  details]{Spera2017, Spera2019, iorio2023}.
Here, we include the envelope binary energy (computed directly from \textsc{\parsec}) as an optional table property in \textsc{sevn}. These can be switched on as an alternative model for the CE treatment. The \textsc{sevn} tables used in this work are publicly available in Zenodo \footnote{\href{https://doi.org/10.5281/zenodo.18340610}{10.5281/zenodo.18340610}}. Here below, we provide a description of the CE phase in \sevn{} and the major changes that our new treatment implies. 
We refer to \cite{iorio2023} for a comprehensive description of the code.

\subsubsection{$(\alpha \lambda)-$formalism}\label{sec:ceformalism}

\sevn{} models the CE phase with the so-called $(\alpha \lambda)-$formalism \citep{denheuvel1976, webbink1984, hurley2002}. This prescription is based on an energy budget criterium: the orbital energy lost due to the shrinkage of the orbit during CE is transferred to the envelope.  This formalism is based on the assumption that the orbital energy constitutes the only energy reservoir able to unbind the envelope. The model depends on two parameters, $\alpha$ and $\lambda$. $\alpha$ represents the efficiency at which energy is transferred from the binary to the envelope. 
Therefore the binding energy of the envelope can be written:
\begin{align}\label{eq:al}
	E_{\rm bind} &= \alpha \Delta E_{\rm orb} \\
	& = \alpha \left( - \frac{G M_{\rm 1} M_{\rm 2}}{2 a_{\rm ini}} + \frac{G M_{\rm core, 1} M_{\rm core, 2}}{2 a_{\rm fin}}\right).
\end{align}
Here, $M$ is the total mass of the star, $M_{\rm core}$ is the core mass of the star and $G$ is the gravitational constant. The subscripts 1 and 2, refer to the primary and secondary stars, respectively. 
$a_{\rm ini}$ and $a_{\rm fin}$ are the pre- and post-CE semi-major axis of the system, required in order to successfully eject the envelope. The value of $a_{\rm fin}$ sets the fate of the CE phase. In \textsc{sevn} if neither of the stars fill their Roche-Lobe after the CE then the envelope is ejected, otherwise the two stars coalesce. 
The second parameter $\lambda$, enters in the definition of the binding energies:
\begin{equation}\label{eq:ebind}
    E_{\rm bind} = E_{\rm bind, 1} + E_{\rm bind, 2}= -G \left( \frac{M_1 M_{\rm env, 1}}{\lambda_1 R_1} + \frac{M_2 M_{\rm env, 2}}{\lambda_2 R_2} \right),
\end{equation}
where $\lambda$ acts as a structural parameter, highly dependent on the stellar properties of the star at the time of CE.
In the previous equations, which we take from \cite{hurley2002}, if the accreting star has not yet developed a core its contribution to the envelope binding energy (Eq. \ref{eq:ebind}) is considered zero, and $M_{\rm core}$ is replaced by its total stellar mass $M$ in Eq.\ref{eq:al}.

The definition of $\alpha$ implies that its value should lie in the range between 0 and 1. Where $\alpha = 1$ corresponds to the ideal scenario where all the missing orbital energy is transferred to the envelope, with no dissipation. In principle, $\alpha > 1$ is still possible if additional energy reservoirs are considered, other than the orbital energy (e.g., recombination energy, outflows, nuclear reactions, \citealt{Ivanova2013a, Giacobbo2018, Fragos2019, Elbadry2023}).

Currently, the default $\lambda$ option in \sevn{} is based on the prescription by \cite{claeys2014}. A constant $\lambda=0.5$ is used for pure-He stars.  Additional $\lambda$ formalisms can be chosen in \sevn{}, based on \cite{Xu2010}, \cite{klencki2021}, or $\lambda$ can be set to a constant value, as in \cite{Spera2019} with $\lambda=0.1$ \citep[see][for more details]{iorio2023}.

\subsubsection{Common Envelope: new features}

We use the envelope binding energies computed directly from the \parsec{} stellar tracks, to evaluate:
\begin{equation}
    \lambda_{\rm i} = - \frac{E_{\rm bind, i} R_{\rm i}}{G M_{\rm i} M_{\rm env, i}},
\end{equation}
where the subscript $i$ refers to the primary (1) or secondary (2) star. We then use Eq. \ref{eq:ebind} to evaluate the total envelope binding energy in a CE phase. For a set $X_{\rm H, 0}$, we compute the envelope binding energies ($E_{\rm bind}$ in the previous equation) through the integrals \ref{eq:eg}, \ref{eq:eb} and \ref{eq:eh}. We store $\lambda$ as an optional table property in \sevn{}. 
In addition, different choices of the core--envelope boundary (i.e., different $X_{\rm H,0}$) affect the definition of the He and CO core masses. Therefore, for any choice of $X_{\rm H,0}$ and $X_{\rm He,0}$, we modify the corresponding tables accordingly. As a consequence, the time of core formation and of the end of the main sequence depends on the adopted $X_{\rm H,0}$, which in turn affects the evolutionary stage at which a star may enter a CE phase (see Section~\ref{sec:results}).

\subsection{Initial conditions} \label{sec:initcond_method}

In order to study the merger rates of binary compact objects, we evolve a population of binaries sampled according to the following initial conditions. 
We draw the primary star masses $M_{\rm ZAMS,1}$ according to \citet{Kroupa2001}, in a range between $5$ and $150\,{} \text{M}_\odot$:
\begin{equation}
\mathcal{F}(M_1) \propto M_{1}^{-2.3}.
\end{equation}
We sample the mass ratios $q=M_{\rm ZAMS,2}/M_{\rm ZAMS,1}$ according to \citep{Sana2012}:
\begin{equation}
\mathcal{F}(q) \propto q^{-0.1},
\end{equation}
with $\max \left( \frac{2.2\,{} \text{M}_\odot}{M_{\rm ZAMS,1}}, 0.1 \right)\leq q \leq 1 $. 
We derive in this way the secondary star mass ($M_{\rm ZAMS,2}$), within the range $[2.2, 150]\,{} {\rm M}_{\odot}$.
The orbital periods and eccentricities are distributed according to \citet{Sana2012}:
\begin{equation}
\mathcal{F}(P_{\rm orb}) \propto (\log P_{\rm orb})^{-0.55},
\end{equation}
with $0.15 \leq \log \left(P_{\rm orb}/{\rm d}\right) \leq 5.5$,  
and
\begin{equation}
\mathcal{F}(e) \propto e^{-0.42},
\end{equation}
with $0\leq e \leq 1-\left(P/2 \ \mathrm{days}\right)^{-2/3}$ \citep{Moe2017}. 
We sample $2\times10^6$ binaries according to these initial conditions. We run this set of binaries for $12$ metallicities: $Z=0.0002$, $0.0004$, $0.0008$, $0.0012$, $0.0016$, $0.002$, $0.004$, $0.006$, $0.008$, $0.012$, $0.016$, $0.02$. The metallicities that are not included among the single stellar tracks are interpolated within \sevn{}, as explained in \citet{iorio2023}.

\subsection{\textsc{sevn} setup} 
We evolve the binaries drawn from the above initial conditions with \textsc{sevn} following the fiducial setup as described by \citet{iorio2023}. We summarize here  the main parameters. 
We adopt the \textit{DelayedGauss} supernova model, which predicts black hole masses according to the delayed supernova model by \citet{Fryer2012}, but draws the masses of neutron stars from a gaussian with mean $1.33$M$_{\odot}$ and standard deviation $0.09$M$_{\odot}$ \citep{ozel2012, ozel2016}. We sample the kicks following \citet{Giacobbo2020}. According to this model, the kicks are drawn from a maxwellian distribution with $\sigma=265$ km s$^{-1}$ \citep{Hobbs2005} and then rescaled by a factor $\propto M_{\rm ej}/M_{\rm rem}$, with $M_{\rm ej}$ the ejected mass and $M_{\rm rem}$ the mass of the remnant. 
We consider the fiducial mass transfer stability criterium (QCRS in \citealt{iorio2023}), that is the same as in \citet{hurley2002}, but donor stars in the main sequence or in the Herzsprung gap lead always to stable mass transfer.
We explore the parameters associated with the CE phase, by varying $\alpha=0.5$, 1, 3 and 5, and the $\lambda$ prescriptions. We compare the models by \citet{claeys2014} and by \citet{klencki2021} with the $\lambda$ parameters evaluated from the \textsc{parsec} stellar tracks, assuming $X_{\rm H,0}=X_{\rm He,0}=10^{-3}$ and $E_{\rm bind} = E_{\rm B}$ (Eq. \ref{eq:eb}). 
We included $\alpha$ values $>1$, to bracket possible uncertainties related to physics that is not captured by the ($\alpha \lambda$)--formalism.

\subsection{Merger rate density} \label{sec:mrd_method}

We evaluate the merger rate density of binary compact objects with \textsc{galaxy}$\mathcal{R}$\textsc{ate} \citep{Santoliquido2022}. 
\textsc{galaxy}$\mathcal{R}$\textsc{ate} adopts data-driven observational scaling relations to generate a set of star-forming galaxies across cosmic time and populate them with binary compact objects. For this work, we use the fiducial setup as described in \citet{Sgalletta2024}. More details on the methodology adopted can be found in Appendix \ref{sec:universe_appendix}.

\section{Results} \label{sec:results}

\subsection{Binding energies}

\begin{figure*}[h!]
   \centering
   \includegraphics[width=0.95\linewidth]{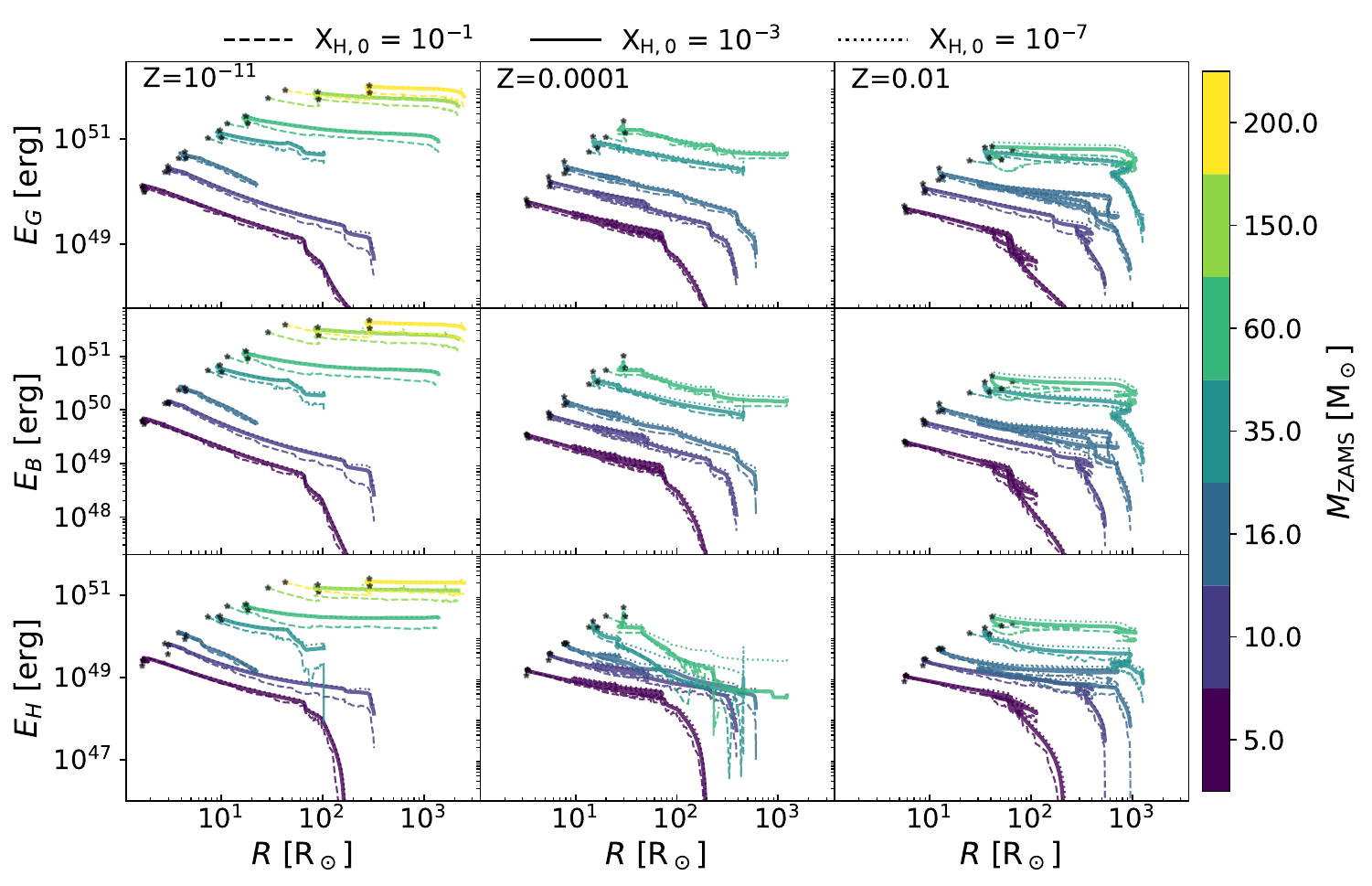}
   \caption{Binding energy $E_H$ evaluated from \textsc{parsec} as a function of the radius $R$ for different stellar masses. The rows from top to bottom show $E_{\rm G}$, $E_{\rm B}$ and $E_{\rm H}$, respectively. Different columns represent different metallicities, from left to right: $Z=10^{-11}$, $0.0001$, and $0.01$. The black star markers indicate the formation of the He core. Different line styles show the results assuming different $X_{\rm H, 0}$ thresholds to define the core-envelope boundaries: $X_{\rm H,0}=10^{-1}$ dashed lines, $X_{\rm H,0}=10^{-3}$ solid lines, and $X_{\rm H,0}=10^{-7}$ dotted lines.
   }
    \label{fig:bindingenergies}%
\end{figure*}

\begin{figure*}
    \centering
   \includegraphics[width=0.95\textwidth]{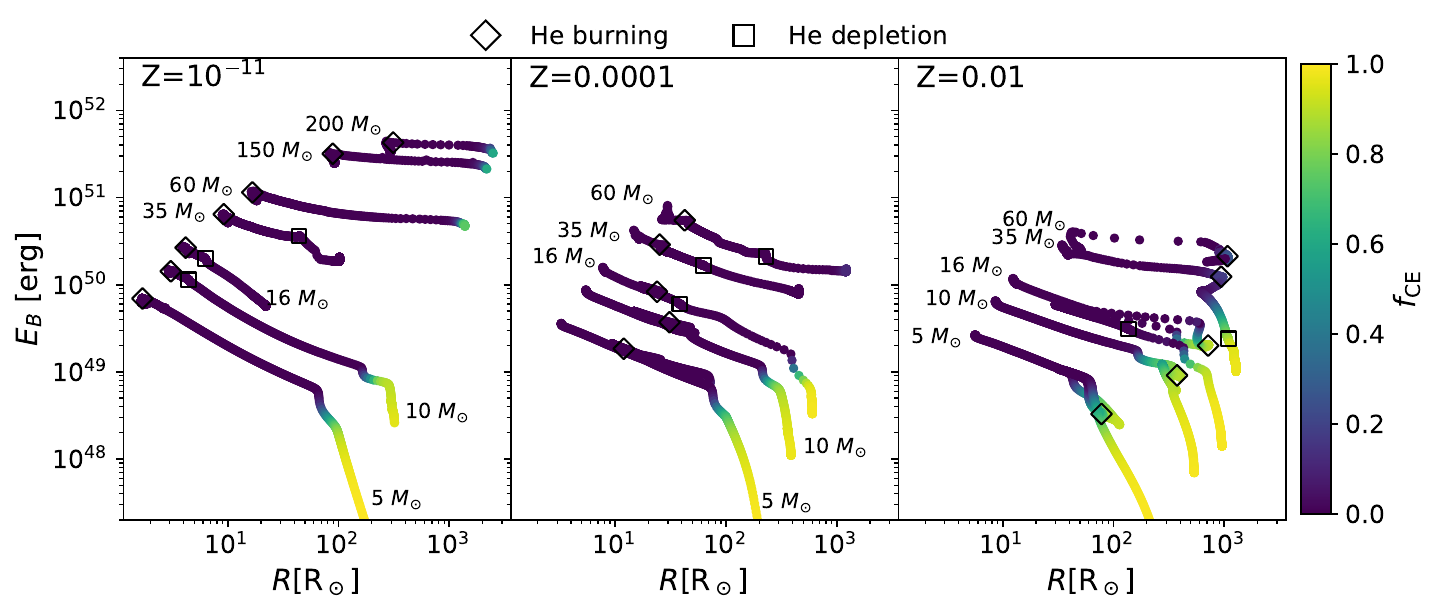}
   \caption{ Binding energies $E_{\rm B}$ evolution with radius for different stellar masses. The color code shows the mass fraction of the outer convective envelope $f_{\rm CE}$. Diamonds indicate the beginning of He burning; squares show the end of the core He burning phase. Different columns represent different metallicities, from left to right: $Z=0.0001$, $0.001$ and $0.01$. Here, we assume $X_{\rm H, 0}=10^{-3}$.}
    \label{fig:convenv}%
\end{figure*}

Figure \ref{fig:bindingenergies} shows $E_{\rm G}$ (Eq.\ref{eq:eg}), $E_{\rm B}$ (Eq.\ref{eq:eb}) and $E_{\rm H}$ (Eq.\ref{eq:eh}) computed for different initial stellar masses and metallicities. 
For visualization purposes, we only show the evolution of massive stars at $Z=0.01$ up to the point where they become Wolf-Rayet stars. This is because, during this phase, stars begin to shrink \citep[e.g.][]{Costa2025}, making them less likely to undergo Roche-lobe overflow. At this metallicity, stars with initial masses $\geq 150 \,{}\mathrm{M}_\odot$ form their cores already during the Wolf-Rayet phase; thus, we omit their evolution from the figure\footnote{The transition to Wolf-Rayet in a massive stellar model depends on the treatment of stellar winds and convection. Therefore, the limits discussed here may vary depending on different prescriptions.}. Low-metallicity stars ($Z \leq 0.0001$) with $M_{\rm ZAMS} \geq 100\,\mathrm{M}_\odot$ typically reach their largest expansion at the end of core helium burning, after which they contract. While the final stellar radii depend on both metallicity and initial mass, they never exceed the maximum radius reached during helium burning \citep[][]{Costa2025}. 
For this reason, we truncate the evolutionary tracks shown in the figures either at maximum expansion ($Z \leq 0.0001$) or when stars become Wolf-Rayet ($Z=0.01$). 
The last point shown corresponds to a stellar age $t=t_{\rm max}$. This evolutionary stage is the most relevant for binary interactions, as the probability of initiating Roche-lobe overflow and a CE phase is maximized near maximum expansion.
Finally, in the case of very massive stars ($M_{\rm ZAMS} \geq 250\,\mathrm{M}_\odot$ at $Z=10^{-11}$ and $M_{\rm ZAMS} \geq 150\,\mathrm{M}_\odot$ at $Z=0.0001$), radial expansion occurs before core formation. Any binary interaction would take place during the expansion phase. Therefore, from a binary population-synthesis perspective, the envelope binding energies of these stars are unlikely to play a major role in typical CE interactions. For clarity, we do not include these stars in the plots.

Generally, smaller core masses exert a lower gravitational pull on the envelope. As a consequence, stars with smaller $M_{\rm ZAMS}$ yield lower binding energies at equivalent radii.
We find differences in the envelope binding energies within a factor of $\sim 2$ among the $X_{\rm H, 0}$ models, for all the stars and metallicities. 
Massive stars at low metallicity ($M_{\rm ZAMS} \geq 150 \mathrm{M}_\odot$, $Z\leq 0.0001$) undergo a significant radial expansion during the early phases of core H burning. Thus, the definition of the helium core is particularly important for the description of mass transfer in such cases. For instance, if we define the helium core as the region where $X_{\rm H, 0}$ falls below 0.1, the star already has a He core during its first Roche-lobe overflow, whereas if we use a more strict definition (e.g.,  $X_{\rm H, 0}=10^{-3}$) the star will be still core-less during mass transfer. On the other hand, the model employing $X_{\rm H, 0} = 10^{-1}$  displays higher fluctuations in the envelope binding energies, suggesting that  the core is still growing because hydrogen burning is still ongoing.

Despite being important, the definition of the He core has a relatively minor impact compared to the definition of envelope binding energy $E_{\rm G}$, $E_{\rm B}$ and $E_{\rm H}$.
Specifically, $E_{\rm G}$ consistently appears to be almost twice as large as $E_{\rm B}$. The difference between $E_{\rm H}$ and $E_{\rm G}$ is even more pronounced, with $E_{\rm H}$ being about one order of magnitude lower than $E_{\rm G}$. Nevertheless, $E_{\rm H}$ is affected by larger fluctuations (see e.g. $M_{\rm ZAMS}=35 \,{}\mathrm{M}_{\odot}$, $Z=0.0001$). 
The reason for this is that the contribution from the enthalpy term progressively approaches the gravitational one as the stars reach the end of their lives. This leads to high fluctuations in $E_{\rm H}$, as the two terms become similar. The effect is particularly evident for $X_{\rm H, 0} = 10^{-1}$. 

Figure \ref{fig:convenv} shows $E_{\rm B}$ as a function of the radius $R$ for several values of the ZAMS mass and metallicity. The color scale indicates the fraction of envelope contained within the outer convective zone:
\begin{equation}
    f_{\rm CE} = \frac{M_{\rm conv}}{M_{\rm env}},
\end{equation}
where $M_{\rm conv}$ is the mass of the outermost contiguous convective zone of the envelope. We can compare Figure \ref{fig:convenv} with Figure 1 in \cite{klencki2021}. 
Our models qualitatively agree with \cite{klencki2021}: we see a drop in the envelope binding energy as the envelope becomes mostly convective ($f_{\rm CE} \geq 0.6$). As observed in \cite{klencki2021}, a deep convective layer does not necessarily mean lower binding energies. Indeed, the massive stars $M_{\rm ZAMS} \geq 60\, \mathrm{M}_\odot$ at $Z=10^{-11}$ and $M_{\rm ZAMS}=150\, \mathrm{M}_\odot$ at $Z=0.0001$, develop a deep convective envelope ($f_{\rm CE} \sim 0.6$), however, they do not show a drop in $E_{\rm B}$. 
In fact, these stars show only a $\sim 10$\% decrease, compared to the $\sim 70 - 90$\% decrease observed in less massive stars when they develope deep convective envelopes.

For massive stars ($M_{\rm ZAMS} \gtrsim 60 \, \mathrm{M}_{\odot} $), the point of core He depletion happens after the maximum radial expansion, therefore it does not appear in the plot.
At $Z = 0.01$, this stage occurs when the star has already become a Wolf-Rayet, while at $Z\leq 0.0001$, helium depletion takes place after the star has reached its maximum expansion radius.

\subsection{The importance of using consistent envelope binding energies in population synthesis}\label{sec:comparison}

\begin{figure*}
    \centering
   \includegraphics[width=0.9\textwidth]{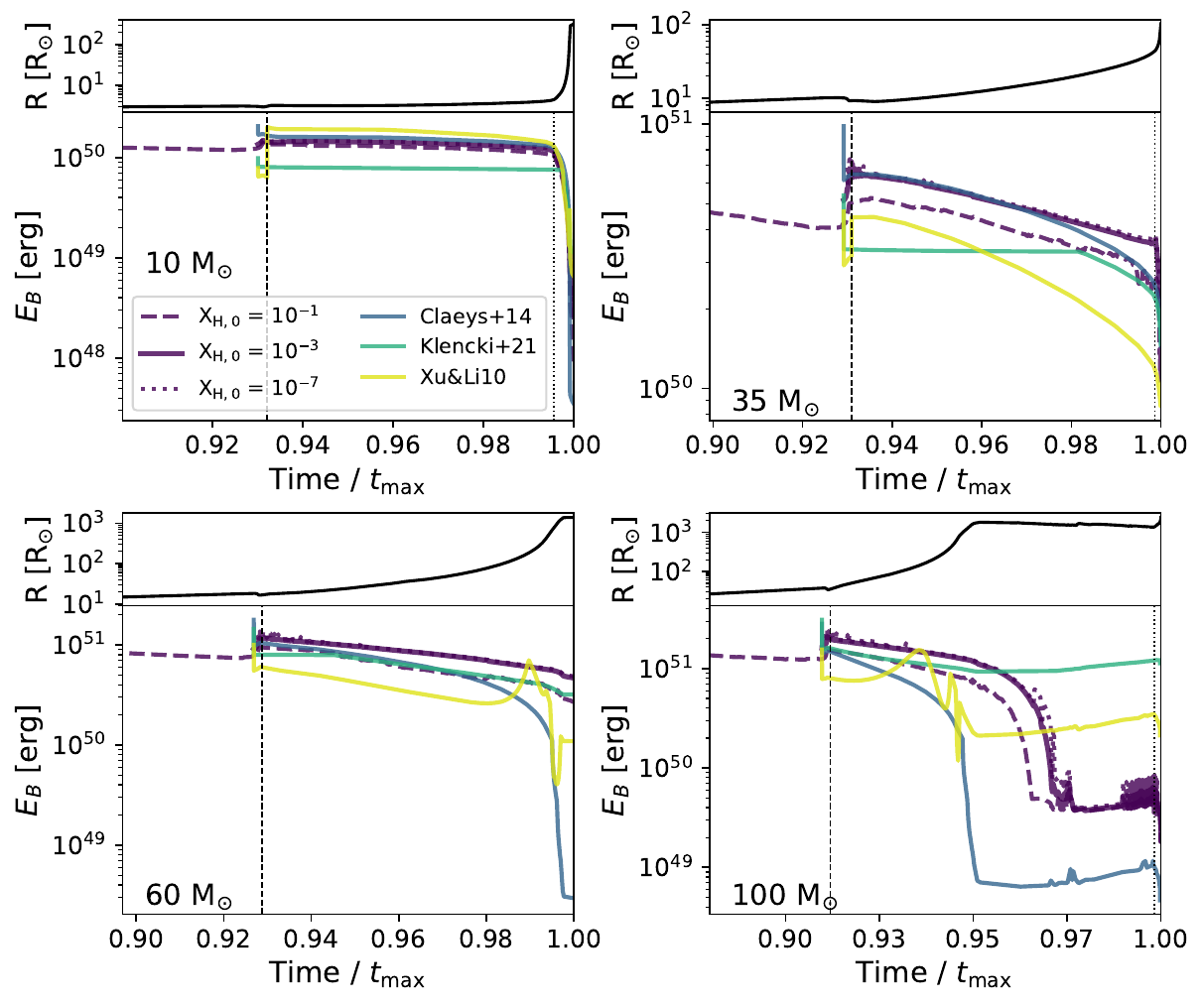}
   \caption{Evolution of the envelope binding energy for a sample of stars with $M_{\rm ZAMS}= 10$ M$_{\odot}$ (top left), $35$ M$_{\odot}$ (top right), $60$ M$_{\odot}$ (bottom left) and $100$ M$_{\odot}$ (bottom right). All the stars have metallicity $Z=10^{-11}$.
   The x-axis is the stellar age normalized to $t_{\rm max}$, where $t_{\rm max}$ is defined as the time at which the star reaches its maximum radial expansion ($Z \leq 0.0001$) or becomes a Wolf-Rayet ($Z = 0.01$). The tracks are truncated at this point in the figures, since after maximum expansion stars typically contract and are unlikely to initiate further common-envelope interactions.
   The purple lines show the envelope binding energies $E_{\rm B}$ computed in this work, for different $X_{\rm H,0}$ thresholds (different linestyles): solid, $X_{\rm H,0}=10^{-3}$; dashed, $X_{\rm H,0}=10^{-1}$; dotted, $X_{\rm H,0}=10^{-7}$. The other colors show different envelope binding energies adapting different $\lambda$ prescriptions to the \textsc{parsec} stellar tracks with \textsc{sevn}: blue solid line assuming \citet{claeys2014}, green solid line \citet{klencki2021}, yellow solid line \citet{Xu2010}. The lines start in different moments in time for the same star due to the different core formation conditions; however, for visualization purposes we do not show the full evolution of the $X_{\rm H,0}=10^{-1}$ model, as it extends well before the other conditions but the star does not evolve much in the meanwhile. The upper panels of each plot show the corresponding evolution of the radii of the stars. The vertical dashed and dotted lines indicate the onset of core He burning and shell He burning, respectively.}
    \label{fig:bindcomp_1e11}
\end{figure*}

\begin{figure*}
    \centering
    \includegraphics[width=0.9\linewidth]{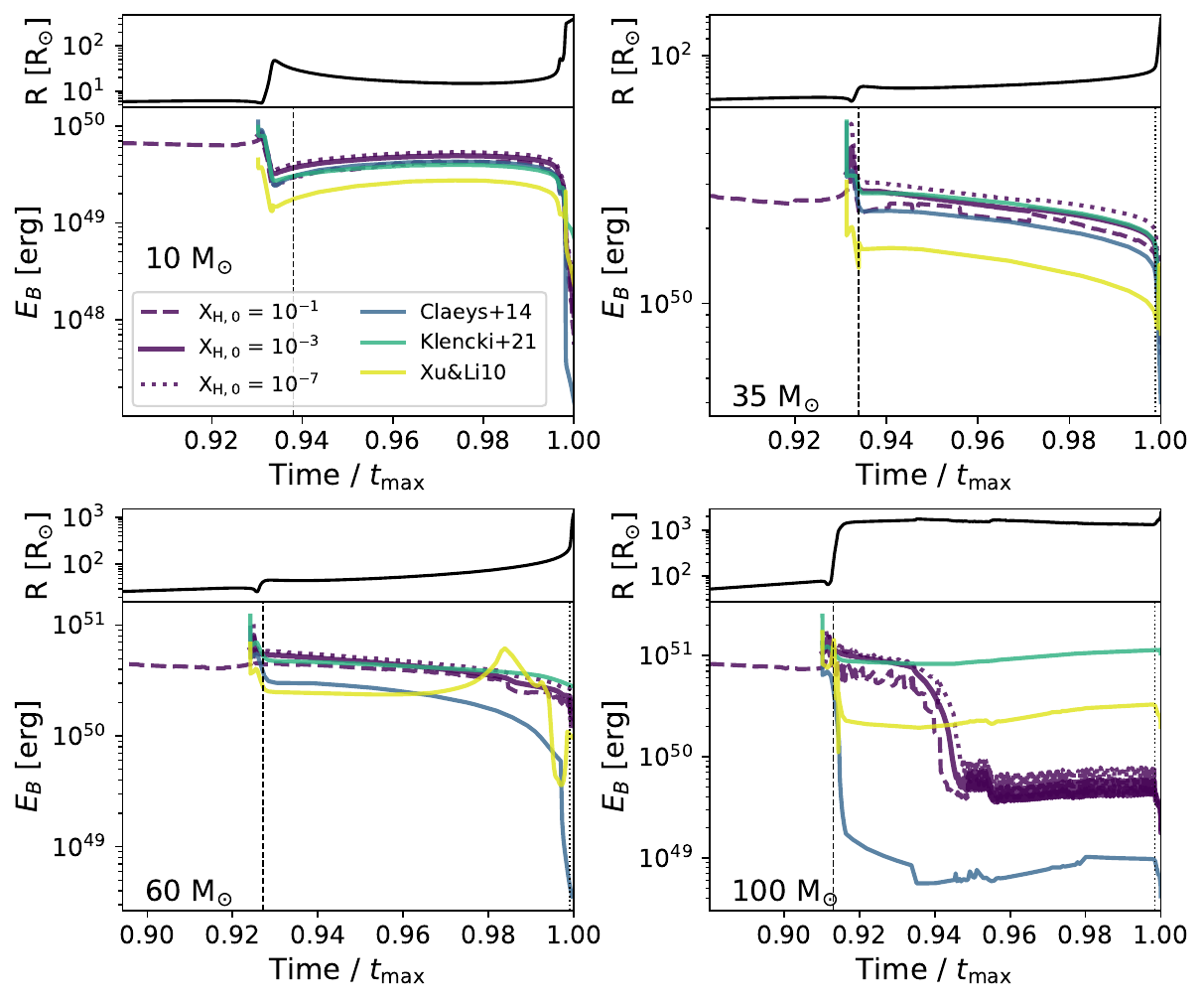}
    \caption{Same as Figure \ref{fig:bindcomp_1e11}, but for metallicity $Z=0.0001$.}
    \label{fig:bindcomp_1e4}
\end{figure*}

\begin{figure*}
    \centering
    \includegraphics[width=0.9\linewidth]{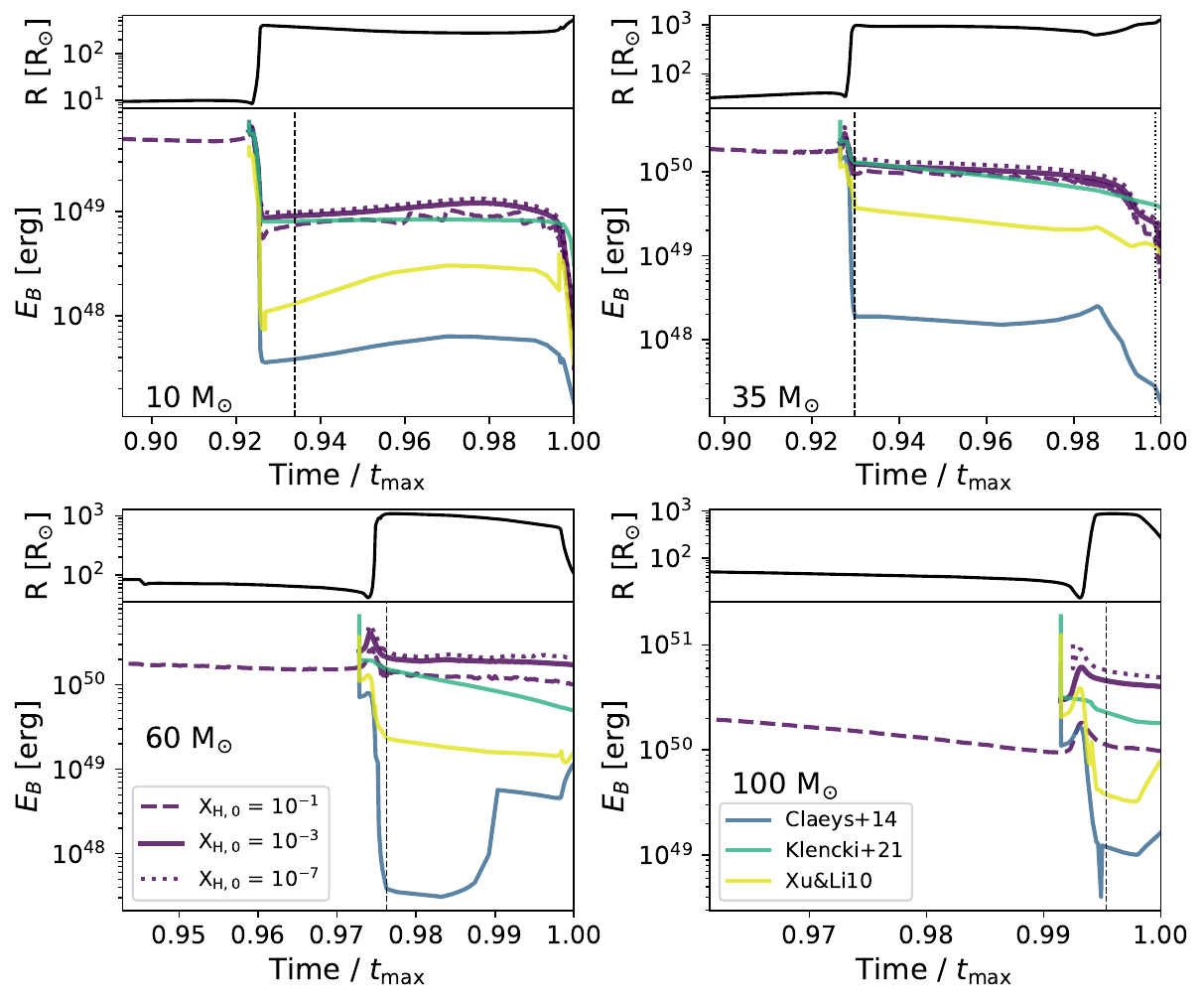}
    \caption{Same as Figure \ref{fig:bindcomp_1e11}, but for metallicity $Z=0.01$.}
    \label{fig:bindcomp_1e2}
\end{figure*}

Figures \ref{fig:bindcomp_1e11}, \ref{fig:bindcomp_1e4} and \ref{fig:bindcomp_1e2} show the envelope binding energy as a function of time, computed with different prescriptions. Among the models (different colors), we observe differences exceeding an order of magnitude. Such differences dominate over variations arising from distinct core-envelope boundary conditions independently of the metallicity and stellar mass.

The model proposed by  \cite{claeys2014}, commonly employed in binary population synthesis codes, consistently underestimates envelope binding energies by more than an order of magnitude as stars approach the giant phase. This is particularly evident at higher metallicities, where the discrepancy can be as high as 3 orders of magnitude (e.g. $M_{\rm ZAMS}=60\,\mathrm{M}_\odot$). Our binding energies are instead in broad agreement with the model by \cite{klencki2021}, obtained with \textsc{mesa}. 
However, extrapolating the model by \citet{klencki2021} outside of the original parameter space ($M_{\rm ZAMS} \leq 80\, \mathrm{M}_\odot$, $Z\leq 0.0001$) leads to large discrepancies with our computed values. This is particularly evident for the star with $Z=10^{-11}$ and $M_{\rm ZAMS}=100\, \mathrm{M}_\odot$. On the other hand, the prescription by \citet{Xu2010}, based on stars with masses between 1 and $100\, \mathrm{M}_\odot$ at $Z=0.001$ and $0.02$, significantly deviates from our model, even considering the same range of stars. 
These substantial divergences highlight the importance of using envelope binding energies that are consistent with the underlying stellar evolution tracks when modeling the CE phase in population synthesis simulations.  

The adoption of different $X_{\rm H, 0}$ affects the core formation times, leading to an earlier formation (i.e., $X_{\rm H, 0}=10^{-1}$) or a delay (i.e., $X_{\rm H, 0}=10^{-7}$).
This effect is likely negligible in binary evolution. In fact in most cases the stars do not undergo significant radial expansion during the period when the $X_{\rm H, 0}=10^{-1}$ model already has formed an He core, while the $X_{\rm H, 0}=10^{-3}$ one has not. 
We note that the envelope binding energies computed in \citet{klencki2021} assume $X_{\rm H, 0}=10^{-1}$, implying that these stars should form the core simultaneously with our dashed line. 
However, when this prescription is applied in \sevn{} with the \parsec{} evolutionary tracks, the core forms instead according to the condition $X_{\rm H, 0} = 10^{-3}$. As a result, the starting point of the model from \citet{klencki2021} coincides with the $X_{\rm H, 0} = 10^{-3}$ model, in the plot.

\subsection{Pure-He stars}

\begin{figure*}
    \centering
    \includegraphics[width=0.95\linewidth]{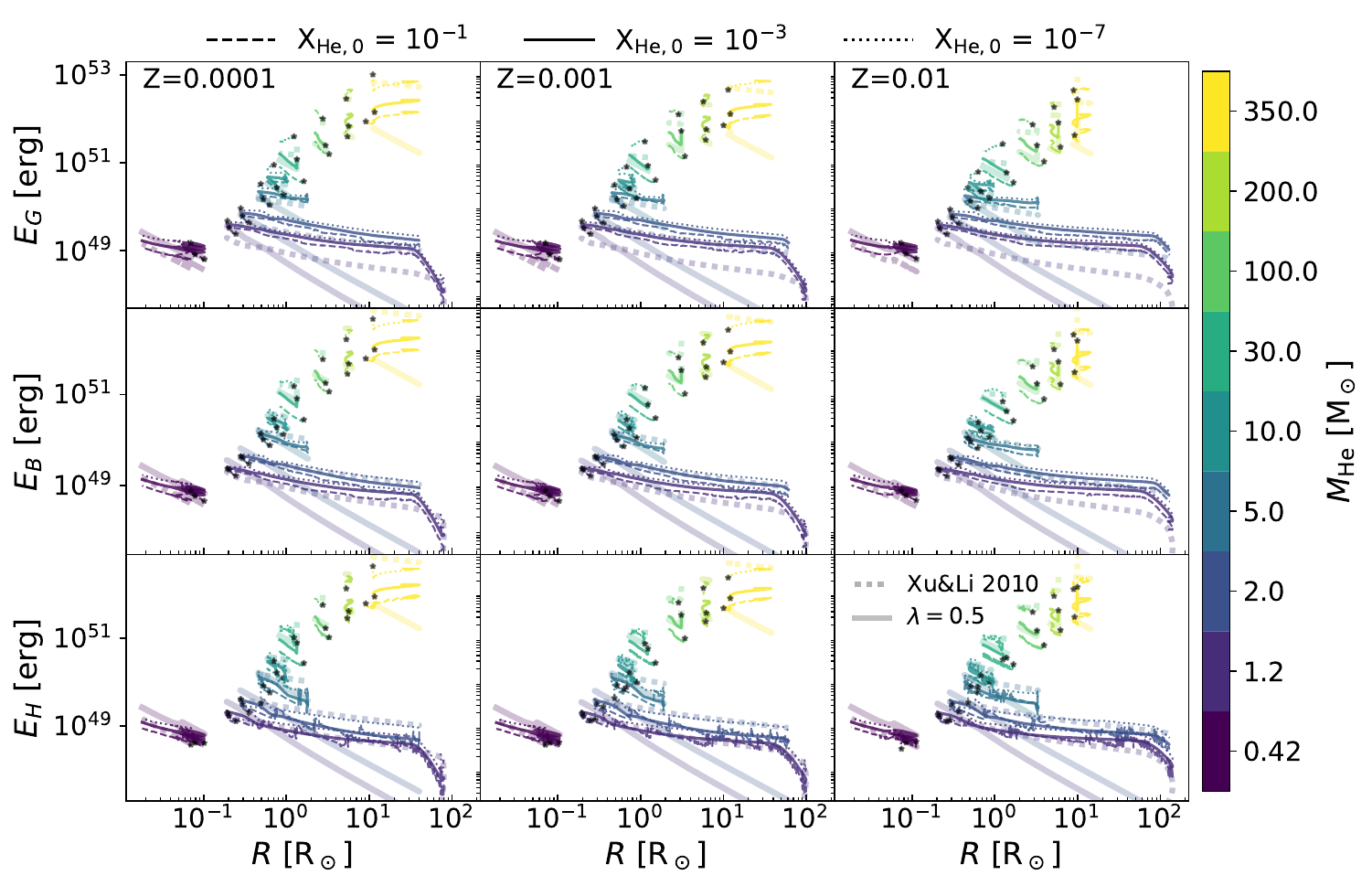}
    \caption{Same as Figure \ref{fig:bindingenergies} but with pure-He tracks. The thick solid lines show the envelope binding energies with constant $\lambda=0.5$, following \citet{claeys2014}. This is the default value adopted in \sevn{} for the prescriptions by \citet{claeys2014} and \citet{klencki2021}.
    The thick dotted lines show the envelope binding energies as implemented in \textsc{compas}. This model is used by \sevn{} with the fitting formulas by \citet{Xu2010}.}
    \label{fig:purehe}
\end{figure*}

Figure \ref{fig:purehe} shows $E_{\rm G}$, $E_{\rm B}$ and $E_{\rm H}$ computed for pure-He tracks with different initial stellar masses and metallicities. For pure-He tracks we show the reference metallicities $Z=0.0001$, $0.001$ and $0.01$. 

Pure-He stars remain significantly more compact throughout their lifetimes compared to hydrogen-rich stars. As a result, also their envelope binding energies span a much narrower range. Moreover, for a given initial mass, the evolutionary tracks are similar across various metallicities. For a wide mass range ( $M_{\rm He} < 1 \,\mathrm{M}_\odot$, $10 < M_{\rm He} < 200\,\mathrm{M}_\odot$), these stars tend to shrink for most of their lives, thus reducing the likelihood of binary interactions.

Unlike H-rich stars, where the contributions to the binding energy differ significantly depending on the considered energy contributions, pure-He stars show much smaller variations, typically no more than a factor of two between $E_{\rm G}$ and $E_{\rm H}$. However, the choice of core-envelope boundary has a substantial impact, especially for high-mass stars ($M_{\rm He} \geq 30\,\mathrm{M}_\odot$).

We also compare our new $E_{\rm bind}$ with two additional models. We show the prescription by \citet{claeys2014}, which assumes a constant $\lambda = 0.5$ for pure-He stars. This is the default value adopted in \sevn{} when using the models by \citet{claeys2014} and \citet{klencki2021}. We also display the fit implemented in \textsc{compas} \citep[][]{riley2022}: 
\begin{equation} \label{eq:compas_fit_ebind}
    \lambda = 0.3 R_{\lambda}^{-0.8} \quad \quad R_{\lambda} = \min (120, \max (0.25, R)),
\end{equation}
where all quantities are in solar units. \sevn{} adopts this fit with the \citet{Xu2010} prescription \citep[see Appendix A1.4 in ][ for more details]{iorio2023}. 

The $\lambda=0.5$ model deviates substantially from the \parsec{} results. This is evident for low-mass stars ($M_{\rm He} \leq 5 \,\mathrm{M}_\odot$), where discrepancies exceed an order of magnitude toward the end of stellar evolution. On the other hand, the envelope binding energies obtained with Eq. \ref{eq:compas_fit_ebind} are in much better agreement with the $E_{\rm bind}$ by \parsec{}. Especially when considering stars with $M_{\rm He} \lesssim 30\,\mathrm{M}_\odot$ and the $E_{\rm B}$ model.

\subsection{Merger rate density}

\begin{figure}
    \centering
    \includegraphics[width=\linewidth]{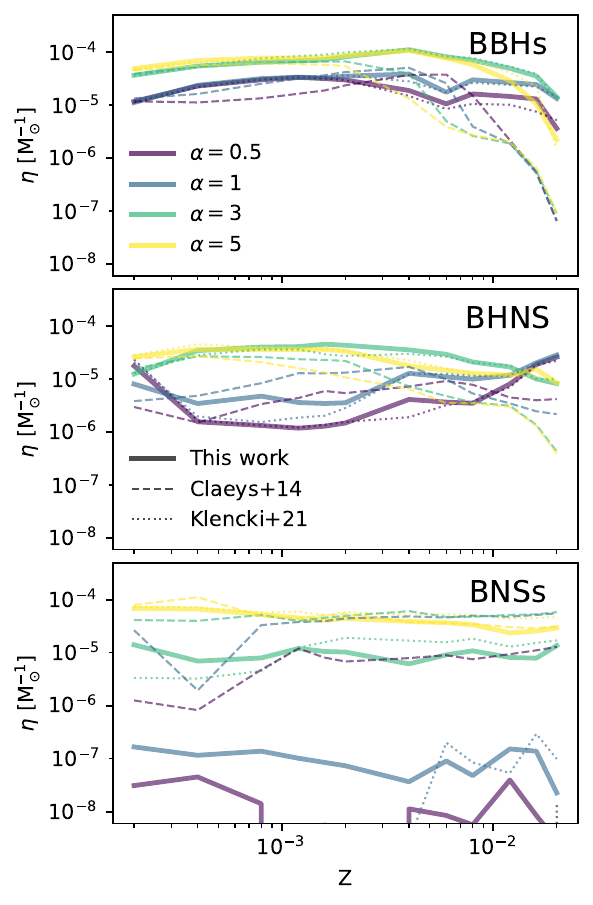}
    \caption{Merger efficiency $\eta$ for BBHs (top), BHNS(middle) and BNSs(bottom) as a function of metallicity $Z$. The solid lines represent the results assuming the \parsec{} envelope binding energies ($E_{\rm B}$) computed with $X_{\rm H,0}=X_{\rm He, 0} = 10^{-3}$. The other linestyles assume different $\lambda$ prescriptions: dashed lines for \citet{claeys2014}, dotted lines for \citet{klencki2021}. 
    The different colors show results for different $\alpha$ parameters: $\alpha=0.5$ (purple), 1 (blue), 3 (green) and 5 (yellow). }
    \label{fig:efficiencies}
\end{figure}

\begin{figure*}
    \centering
    \includegraphics[width=\textwidth]{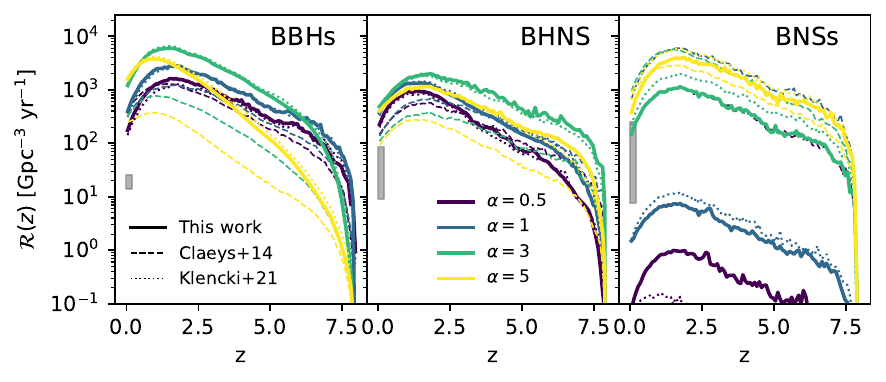}
   \caption{Merger rate density of BBHs (left), BHNS (center) and BNSs (right) for different $\lambda$ prescriptions and $\alpha$ parameters. The solid lines show the results adopting the $\lambda$ self-consistent with the \textsc{parsec} tracks, derived from $E_{\rm B}$ and $X_{\rm H,0}=X_{\rm He, 0} = 10^{-3}$. The dashed lines assume the $\lambda$ prescription by \cite{claeys2014}, the dotted lines assume the prescription by \citet{klencki2021}. The different colors show the different $\alpha$ parameters: purple $\alpha=0.5$, blue $\alpha=1$, green $\alpha=3$ and yellow $\alpha=5$. The grey shaded areas show the merger rate densities inferred by LVK \citep{lvk2025_population}. }
    \label{fig:mrds}%
\end{figure*}

We estimate the merger rate density of binary compact objects following the methodology outlined in Section \ref{sec:mrd_method}. Figure \ref{fig:mrds} shows the resulting merger rate densities for binary black holes (BBHs, left), black hole--neutron star (BHNS, middle) and binary neutron stars (BNSs, right) with our new consistent envelope binding energies (solid lines). 
The merger rates' differences between our new formalism and \citet{claeys2014} are as high as a factor of two for many of the models.
The effect of the new $E_{\rm bind}$ is to boost the merger rate densities of BBH and BHNS systems for $\alpha \gtrsim 1$. Conversely, the new values of $E_{\rm bind}$ quench the BNS merger rate densities, with a more apparent effect for $\alpha\leq 1$.

The merger rate densities agree well with the model by \citet{klencki2021}. 
This was expected, given the similarities among the two $E_{\rm bind}$ prescriptions. 

Differences in the BBH merger rate densities are primarily driven by variations of the merger efficiency's dependence on metallicity, as illustrated in Figure \ref{fig:efficiencies}. For $\alpha=3$,~5 the envelope binding energies by \citet{claeys2014} predict a drop in the BBH merger efficiency for $Z \sim 5 \times 10^{-3}$. In contrast, with the \textsc{parsec} envelope binding energies, the merger efficiency stays constant up to $Z \sim 10^{-2}$. The latter trend is very similar to the model by \citet{klencki2021}, where, as showed in \citet{iorio2023}, the drop occurs only for $Z\gtrsim 10^{-2}$. 
Since higher metallicities play a major role in the cosmic star formation history \citep[][]{Sgalletta2024}, both the \parsec{} and the \citet{klencki2021} $E_{\rm bind}$ predict a boost in the BBH merger rate density.
On the other hand, for $\alpha < 1$ the BBH merger efficiency is lower than the one obtained adopting the model by \citet{claeys2014}, considering metallicities $2\times 10^{-3} \lesssim Z \lesssim 10^{-2}$. Similarly, the merger rate density of BNSs for $\alpha=0.5$,$1$ is significantly suppressed as a consequence of lower merger efficiencies.

The envelope binding energies by \citet{claeys2014} are consistently lower than those predicted by \textsc{parsec}. Higher binding energies have two primary effects: making the shrinkage during CE more efficient, thus leading to lower post-CE separations, and increasing the occurrence of failed CE ejections. For BBH systems at high metallicity (i.e., $Z=0.02$), the former effect dominates. While most BBHs remain wider in \citet{claeys2014}, the new envelope binding energies facilitate orbital shrinkage. This leads to a substantial population of systems being able to merge within a Hubble time even at Z $\sim$0.02. Conversely, for BNSs  and low values of $\alpha$, failed CE ejections constitute the dominant effect.

\section{Discussion} \label{sec:discussion}

\begin{figure*}
    \centering
    \includegraphics[width=0.95\linewidth]{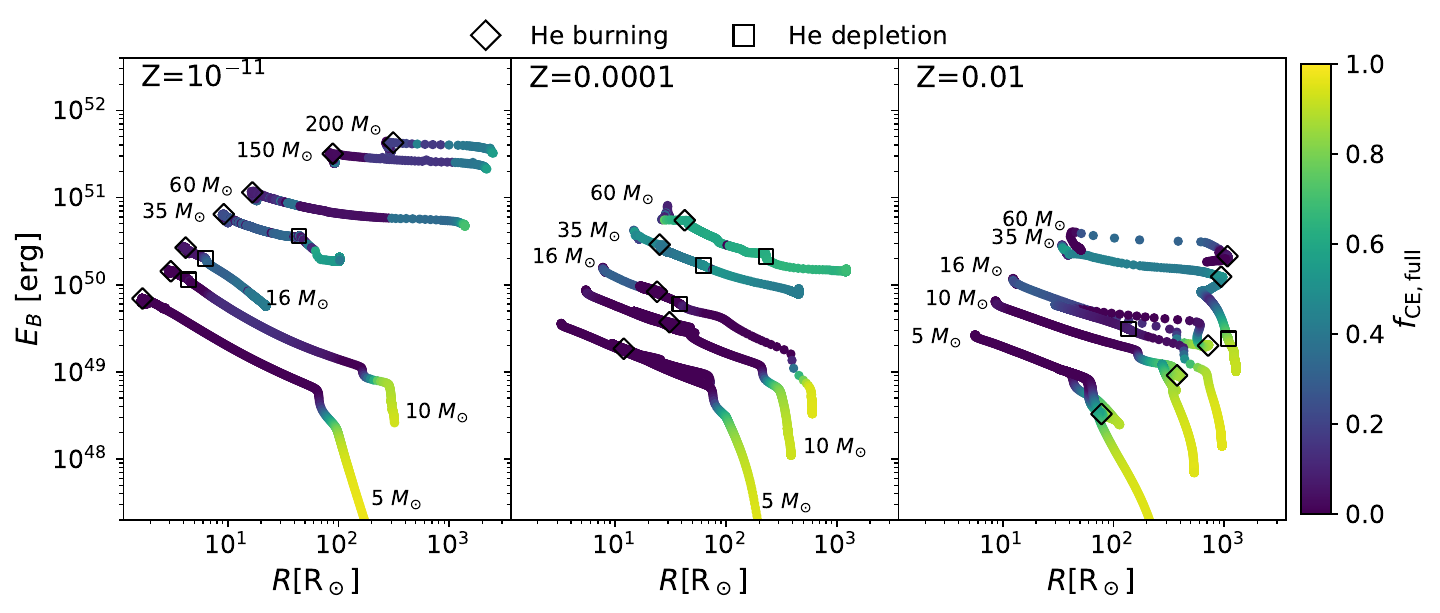}
    \caption{Same as Figure \ref{fig:convenv}, but colored by $f_{\rm CE, full}$, the fraction of the total convective envelope mass over the envelope mass.}
    \label{fig:convenv_full}
\end{figure*}

\citet{Sgalletta2024} has pointed out that current binary population synthesis codes tend to predict higher BBH merger rate densities compared to inferred rates from the LIGO-Virgo-KAGRA collaboration \citep{Abbott2023population, lvk2025_gwtc4}. The formalism presented here accentuates this discrepancy, pushing the BBH and BHNS merger rates well outside the 90\% credible intervals if $\alpha > 1$. In order to relieve the tension we might therefore favor models with $\alpha < 1$. 
However, the models with $\alpha < 1$ produce BNS merger rates below the 90\% credible interval inferred from observations \citep{lvk2025_population}.
A possible solution would be a varying $\alpha$ parameter, depending on the properties of the stars within the CE phase. Studies on double white dwarf systems have already highlighted the possibility of  $\alpha$ being a function of the masses of the white dwarfs, as demonstrated by \cite{Ivanova2013b, roepke2023}. Similarly, \citet{chruslinska2018} proposed different CE ejection mechanisms to explain the rates of BNSs and BBHs. Our results support the scenario of a varying $\alpha$ parameter for massive progenitor stars. The recently developed two-stage formalism for CE treatment \citep{Hirai2022} finds evidence for a broad range of $\alpha$ parameters, opposed to a universal value \citep{Picker2024}. 

Nevertheless, several caveats remain. Our analysis is based purely on single stellar evolutionary tracks. A detailed response of stars to mass transfer and mass accretion could be crucial in understanding the final fate of these systems. Indeed, recent studies suggest that the CE channel might occur less frequently than currently assumed in population synthesis calculations \citep[e.g.,][]{Ge2020, Gallegos2021, Marchant2021, VanSon2022}.
An additional caveat concerns the fact that envelope binding energies are computed assuming envelopes bound to single stars in hydrostatic equilibrium, whereas during the CE phase the envelope becomes part of a highly dynamical binary system, and its initial binding energy may not be conserved. As a result, the envelope binding energy governing the interaction may differ from that inferred from the initial stellar configuration.

A final comment regards the role played by convection in affecting the fate of mass transfer events \citep{klencki2021, Hirai2022}. Figure \ref{fig:convenv_full} is the same as \ref{fig:convenv}, but the colors show:
\begin{equation}
    f_{\rm CE, full} = \frac{M_{\rm conv, full}}{M_{\rm env}},
\end{equation}
where $M_{\rm conv, full}$ is the total mass of the convective envelope, accounting  for both the outer convective zone and the inner convective zones within the stellar envelope. Compared to Figure \ref{fig:convenv}, we observe that stars (especially those with masses $\geq 30$ M$_\odot$) develop internal convective regions even before these zones reach the surface of the star. In these cases, the response of the star is thus unclear. There may be episodes in which the radiative layers are stripped away and the response of the star becomes the same as for a convective star. A model that considers only the outer convective layers fails to capture this complexity.

\section{Summary} \label{sec:summary}

We have derived the envelope binding energies for an extensive set of stars  in terms of both masses and metallicities. 
We have investigated different energy contributions to the envelope binding energy as well as alternative core-boundary criteria. Additionally, we studied how different prescriptions influence the predicted merger rate densities of compact object systems. We summarize our main findings  below.

\begin{itemize}
    \item In the case of H-rich stars, varying the $X_{\rm H,0}$ threshold affects the envelope binding energies by no more than a factor of $\sim 2$ in most cases. However, the choice of which energy sources to include ($E_{\rm G}$, $E_{\rm B}$ or $E_{\rm H}$) has a more significant impact, with $E_{\rm H}$ being roughly an order of magnitude smaller than $E_{\rm G}$. Conversely, for pure-He stars, the choice of $X_{\rm He,0}$ has a greater influence on the envelope binding energy than the selected energy contributions.

    \item  The inconsistencies among various envelope binding energy prescriptions, each evaluated with different stellar evolution codes, emphasize the need to adopt binding energies that are consistent with the specific stellar models being used. Additionally, it is crucial to avoid extrapolating these prescriptions beyond the parameter space for which they were originally inferred.

    \item Differences in envelope binding energy models directly translate into differences in the predicted merger rate densities, with discrepancies exceeding an order of magnitude.
     Compared to the prescription by \citet{claeys2014}, the higher \parsec{} envelope binding energies cause a boost in the merger rate densities of BBHs and BHNS for $\alpha \geq 1$. Concurrently, they suppress the BNS merger rate density for $\alpha < 1$. This trend is in agreement with the model by \citet{klencki2021}, which also predicts similar envelope binding energies.
\end{itemize}

\begin{acknowledgements}
The authors thank the anonymous referee for the constructive report. GI thanks Jakub Klencki for his support and collaboration on debugging and improving the implementation of the Klencki+21 binding-energy prescription in SEVN (version > 2.10.0, see \url{https://gitlab.com/sevncodes/sevn/-/issues/4}). 
CS acknowledges financial support from the Alexander von Humboldt Foundation for the Humboldt Research Fellowship. MS acknowledges financial support from Large Grant INAF 2024 ``Envisioning Tomorrow: prospects and challenges for multi-messenger astronomy in the era of Rubin and Einstein Telescope'', from Fondazione ICSC, Spoke 3 Astrophysics and Cosmos Observations, National Recovery and Resilience Plan (Piano Nazionale di Ripresa e Resilienza, PNRR) Project ID CN\_00000013 ``Italian Research Center on High-Performance Computing, Big Data and Quantum Computing'' funded by MUR Missione 4 Componente 2 Investimento 1.4: Potenziamento strutture di ricerca e creazione di ``campioni nazionali di R\&S (M4C2-19 )'' - Next Generation EU (NGEU), and from the program ``Data Science methods for Multi-Messenger Astrophysics \& Multi-Survey Cosmology'' funded by the Italian Ministry of University and Research, Programmazione triennale 2021/2023 (DM n.2503 dd. 09/12/2019), Programma Congiunto Scuole.
MM acknowledges financial support from the European Research Council for the ERC Consolidator grant DEMOBLACK, under contract no. 770017. CS and MM acknowledge financial support from the German Excellence Strategy via the Heidelberg Cluster of Excellence (EXC 2181 - 390900948) STRUCTURES.
GC acknowledges financial support from European Union—Next Generation EU, Mission 4, Component 2, CUP: C93C24004920006, project ‘FIRES’.
GI is supported by a fellowship grant from la Caixa Foundation (ID 100010434). The fellowship code is LCF/BQ/PI24/12040020.
We use the \textsc{sevn} version V 2.16 (commit \href{https://gitlab.com/sevncodes/sevn/-/tree/8af02cc3706b132609cb3b7fb3ca029c8629bfb7}{8af02cc3}) to generate our binary compact objects catalogs. \textsc{sevn} is publicly available at the gitlab repository \url{https://gitlab.com/sevncodes/sevn}. We use \textsc{trackcruncher} (\url{https://gitlab.com/sevncodes/trackcruncher}) \citep{iorio2023} to produce the tables needed for the interpolation in \textsc{sevn}. We estimate the binary compact objects merger rate densities using the code \textsc{galaxy$\mathcal{R}$ate}. \textsc{galaxy$\mathcal{R}$ate} can be found at \url{https://gitlab.com/Filippo.santoliquido/galaxy_rate_open}.
This research made use of \textsc{NumPy} \citep{Harris20}, \textsc{SciPy} \citep{SciPy2020}, \textsc{Pandas} \citep{Pandas2024} and \textsc{Astropy} \citep{astropy:2013, astropy:2018, astropy:2022}. For the plots we used \textsc{Matplotlib} \citep{Hunter2007}. 
\end{acknowledgements}

\bibliographystyle{aa}
\bibliography{ref}

\begin{appendix}

\section{Our synthetic galaxy model} \label{sec:universe_appendix}

Here, we summarize the details of the  model  we used to calculate binary compact objects merger rate densities. We refer to the fiducial model described by \citet{Sgalletta2024} for additional details.

\subsection{The galaxy stellar mass function}
We simulate a comoving volume $V\sim (70$ cMpc$)^3$ and populate it with galaxy stellar masses $M_\star$ drawn from the galaxy stellar mass function derived by \citet{Chruslinska2019}. The fit takes the following form: 
\begin{equation}
    \phi(M_\ast, z) \,{}dM_\ast = \phi_{\rm N}(z) \,{} e^{-M_\ast / M_{\rm cut}(z)} \left( \frac{M_\ast}{M_{\rm cut}(z)}\right)^{-\alpha_{\rm GSMF}} dM_\ast,
\end{equation}
where $\phi_{\rm N}(z)$ is a normalization factor and $M_{\rm cut}(z)$ is the galaxy stellar mass at which the function transitions from a power law at low masses to an  exponential decay at high masses.
We sample the galaxy masses within the range $M_\star \in [10^{7}, 10^{12}]\, \mathrm{M}_\odot$. We generate in this way $\mathcal{N}_{\rm gal} \left( z \right)$ for an array of redshifts, from 0 to $z_{\rm max}=8$. 

\subsection{Galaxy mass -- star formation rate}

We assume the galaxy main sequence relation derived in \citet{Popesso2023} to determine the galaxies' star formation rates.
We draw the galaxies' star formation rates from a double log-normal distribution\citep{Sargent2012, Rodighiero2015, Schreiber2015}:
\begin{multline}
    \mathcal{P} (\log{\rm SFR} | M_\ast, z) = A_{\rm MS} \exp\left[  - \frac{\left( \log{\rm SFR} - \langle \log{\rm SFR} \rangle_{\rm MS} \right)^{2}}{ 2 \sigma_{\rm MS}^{2}}\right] +\\  A_{\rm SB} \exp\left[  - \frac{\left( \log{\rm SFR} - \langle \log{\rm SFR} \rangle_{\rm SB} \right)^{2}}{ 2 \sigma_{\rm SB}^{2}}\right],
\end{multline}
where $A_{\rm MS}=0.97$ and $A_{\rm SB} = 0.243$ constants \citep{Sargent2012}. Here, the subscript MS (SB) stands for galaxy main sequence (starburst sequence). 
We assume the galaxy main sequence to follow the relation by \citet{Popesso2023}:
\begin{equation} \label{eq:p23}
    \langle \log \mathrm{SFR} \left( M_\ast, t\right) \rangle_{\rm MS} = a_0 + a_1 t - \log \left[ 1 + \left( M_\ast / 10^{a_2 + a_3 t}\right)^{-a_4} \right]
\end{equation}
with $a_0 = 2.693 \pm 0.012$, $a_1=-0.186 \pm 0.009$, $a_2=10.85 \pm 0.05$, $a_3=-0.0729 \pm 0.0024$ and $a_4=0.99 \pm 0.01$.
On the other hand, the starburst sequence parameters are defined as in \cite{Sargent2012},
\begin{equation}
    \langle \log{\rm SFR} \rangle_{\rm SB} = \langle \log{\rm SFR} \rangle_{\rm MS} + 0.59 
\end{equation}
We assume $\sigma_{SB} = 0.243$ dex.

\subsection{Fundamental metallicity relation}

Finally, we associate an avarage metallicity to each galaxy according to the fundamental metallicity relation by \citet{Andrews2013}: 
\begin{equation}
    12 + \log \left(\mathrm{O/H} \right) = 0.43 ( \log M_\ast - 0.66 \log{\rm SFR} ) + 4.59.
\end{equation}

Within each galaxy, we distribute the metallicities following a log-normal function around this average value, with a scatter $\sigma_{\rm Z}=0.14$.
We adopt the solar metallicity values from \cite{Caffau2011}: $Z_{\odot} = 0.0153$ and $12 + \log (\mathrm{O/H})_{\odot} = 8.76$ to convert the relation into absolute metallicity.  

\subsection{Merger rate density evaluation}

We estimate the merger rate density as \citet{Santoliquido2022}:
\begin{equation} \label{eq:mrd}
    \mathcal{R}(z) = \frac{1}{V^3} \int_{z_{\rm max}}^{z} 
    \left[ \int_{Z_{\rm min}}^{Z_{\rm max}} \mathcal{S} \left( z', Z \right) \mathcal{F}\left( z', z, Z \right) dZ \right] \frac{dt(z')}{dz'} dz',
\end{equation}
where $V$ is the total simulated comoving volume. 
The star formation rate density $\mathcal{S}(z', Z)$ is given by:
\begin{equation}
     \mathcal{S}(z', Z) = \sum_{ \rm{i} = 1}^{\mathcal{N}_{\rm gal}(z')} \psi_{\rm i} \left( z'\right) p_{\rm i}\left( Z \vert z' \right)
\end{equation}
where we sum over all the galaxies $\mathcal{N}_{\rm gal}(z')$ formed at redshift $z'$. For the i-th galaxy, $\psi_{\rm i} \left( z'\right)$ is its star formation rate and $p_{\rm i}\left( Z \vert z' \right)$ its metallicity distribution. 
The term $\mathcal{F}\left( z', z, Z \right)$ is defined as:
\begin{equation} \label{eq:catalogs}
    \mathcal{F}\left( z', z, Z \right) = \frac{1}{M_{\rm sim}} \frac{\mathcal{N} \left( z', z, Z \right)}{dt} f_{\rm bin} f_{\rm corr},
\end{equation}
where $M_{\rm sim}$ is the total simulated initial mass, $\frac{\mathcal{N} \left( z', z, Z \right)}{dt}$ is the rate of binary compact objects that form at $z'$, with metallicity $Z$ and merge at redshift $z$. The factor $f_{\rm corr} =0.251$ takes into account the incomplete sampling of the initial mass function (see Section \ref{sec:initcond_method}), $f_{\rm bin}=0.5$ \citep{Moe2017} represents the fraction of binaries.

\end{appendix}

\end{document}